\documentclass[11pt]{article}

\usepackage{cospar}
\usepackage{psfig}
\usepackage{url}
\usepackage{graphicx}
\input psfig.sty

\begin{document}

\title{THE PECULIARITIES OF THE INTERACTION OF PHOBOS WITH THE SOLAR
WIND ARE EVIDENCE OF THE PHOBOS MAGNETIC OBSTACLE \\
(FROM PHOBOS-2 DATA)}

\author{V.G. Mordovskaya$^{1}$, V.N. Oraevsky$^{1}$, and V.A.
Styashkin\address{Institute of Terrestrial Magnetism, Ionosphere, and
Radiowave Propagation (IZMIRAN), Troitsk, Moscow region, 142190 Russia,
mail: valen@izmiran.rssi.ru}}

\maketitle

\begin{abstract}
Using magnetic field and plasma data acquired during {\it Phobos-2}
mission in regions which have not been explored before, we study the
solar wind interaction with Phobos. The draping magnetic field of the
solar wind around Phobos appears at distances of 200--300~km from the
Phobos day-side due to a density and magnetic field pile up in front of
the Phobos obstacle. The nature of the interaction and the magnetic
field signatures observed are consistent with the ratio of the proton
skin depth to the actual size of the Phobos obstacle to the solar wind.
Phobos deflects the flow of the solar wind and the subsolar stand-off
distance of the deflection is about 16--17 Phobos radii. Source with
equivalent magnetic moment $M'\simeq10^{15}$ A$\cdot$m$^2$ in Phobos
leads to the development of such an obstacle to solar wind flow around
Phobos. These data give a lesson for the study of the interaction of
the small, magnetized object with solar wind.
\end{abstract}

\section*{INTRODUCTION}

The last planetary explorations have witnessed an increase of
importance of investigations of small bodies of the Solar System
because their substance may carry the information on an origin and
evolution of Solar system. Interplanetary spacecraft have passed near
several asteroidal bodies, i.e., Gaspra (20 km, 11 km), Ida (56 km, 21
km), Braille (1 km, 2.2 km), Eros (14 km, 34 km), and Martian moons
Phobos (9.5 km, 13.5 km) and Deimos (5.5 km, 7.5 km). Within the
parentheses given are the shortest and longest sizes of the bodies.

Galileo data showed the existence of magnetic field rotations near
asteroids Gaspra at the closest approach distance 1600 km and Ida at
2700 km which have been interpreted as signatures of the solar wind
interaction with the asteroids (Kivelson et al., 1993, 1995).
Blanco-Cano et al. (2002) from own simulation results led to conclusion
that the perturbation near Gaspra was not generated by the interaction
of the solar wind with a magnetized asteroid and the signature observed
near Ida was not generated by the interaction with the asteroid.

A 1 to 2 nT increase in the magnetic field perturbation at the distance
28 km from Braille was attributed by Richter et al. (2001) to a direct
measurement of Braille magnetic field.

The solar wind interaction with the asteroid 433 Eros was investigated by
the use of magnetometer data from the Near-Shoemaker spacecraft having
circular orbits with radii of 200 km, 100 km, 50 km, and 35 km
(Anderson and Acuna, 2003). No signature of an intrinsic
asteroidal magnetic field was detected for Eros. Any draping or
compressional signature associated with Eros was absent. Eros and
Martian moons, Phobos and Deimos, have the same order of the ratio of
the ion gyroradius and the ion skin depth to the body size. The
magnetic field signatures of their interaction with the solar wind
plasma have to be similar if the moons can be considered as obstacles
having the size of body.

However, the analysis of the magnetic field and plasma data aboard the
Mars-5 spacecraft indicated that the Deimos obstacle to the solar wind
is about 100 km (Bogdanov, 1977). It is much larger then the Deimos
size.

The trajectory of {\it Phobos}-2 provided the collection of data in
regions which are appropriate for an investigation of interaction of
Phobos with the solar wind and have not been explored before. The study
of the interaction of Phobos with the solar wind plasma indicated that
the day-side obstacle of Phobos to the solar wind is over 150 km. A
sharp rise in the regular part of the magnetic field was observed on
the circular orbits near Phobos at distances of 180--250~km from its
center when Phobos was in the unperturbed solar wind (Yeroshenko,
2000).

Analyzing the data acquired aboard {\it Phobos}-2, Mordovskaya et al.
(2001, 2002) gave evidence that Phobos has its own magnetic field and
its magnetic moment is $M'\simeq10^{15}$ A$\cdot$m$^2$. The magnetic
moment $M'$ was estimated from pressure balance for the solar wind and
the Phobos magnetic field measured at the magnetopause. The peculiarity
of the rotation of the magnetized Phobos around Mars leads to the
magnetic field signatures, which, especially the direction,
are phase locked with Phobos rotation rate. Such magnetic field
signatures were observed on circular orbit of the Phobos-2 spacecraft
(Mordovskaya et al., 2002).

In the present paper, we investigate the magnetic field enhancement
observed near the dayside of Phobos in order to show that it is the
kinetic theory that should be used when studying the solar wind
interaction with a small magnetized bodies. We consider a density and
magnetic field pile up in front of Phobos to illustrate that the
magnetic field signatures observed are consistent with the ratio of the
proton skin depth to the actual size of the Phobos obstacle to the
solar wind. The occurrence or the absence of the shock-like structures
ahead of the Phobos magnetopause correlates with behaviour of this
ratio. In addition, we demonstrate how the solar wind is stopped and
deflected away from Phobos by a magnetic barrier, which is compressed
or expands depending on the solar wind ram pressure variations.

\section*{THE MORPHOLOGY OF THE MAGNETIC FIELD SIGNATURES NEAR PHOBOS
ON CIRCULAR ORBITS FROM MARCH 22 UNTIL MARCH 26, 1989}

\subsection*{Data Set}

\begin{minipage}{85mm}
\rightskip 5mm
\vskip 10mm
\psfig {file=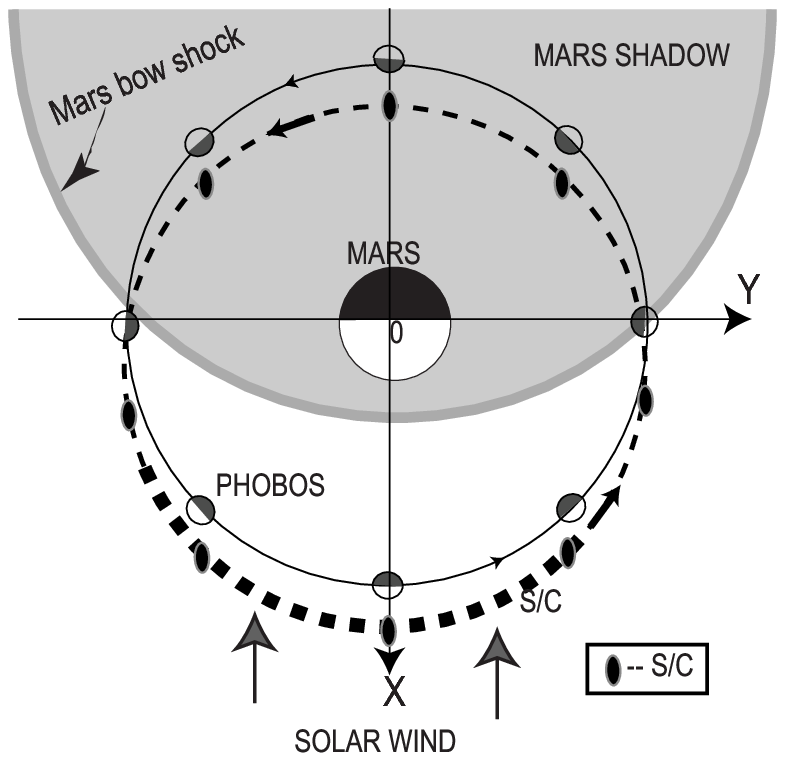}
{\small {\bf Fig. 1.}  {\bf V}iew of the location of
the {\it Phobos}-2 and Phobos on March 22--26, 1989. Plot by line
composed of small squares shows the segment of the S/C trajectory
where the field enhancements were observed.}
\end{minipage}
\begin{minipage}{90mm}
\vspace{3mm}
Vector measurements of the ambient magnetic field
during {\it Phobos}-2 mission were acquired by two magnetometers,
FGMM and MAGMA (Riedler et al., 1989). Both instruments had a
dynamic range of 100 nT and data were transmitted every 1.5, 2.5,
45, and 600 s, depending on the telemetry mode of the spacecraft.
From March 22, 1989, to March 26, 1989, at each orbit around Mars,
both {\it Phobos}-2 spacecraft and the Mars satellite Phobos were
inside the solar wind and within the Martian magnetosphere during
3.8 h. The spacecraft was located permanently in a vicinity of
Phobos at this time and the distances between them were 180--400
km. A description of the spacecraft flight profile is given by
Kolyuka et al. (1991). Figure~1 displays a sketch of the position
of the {\it Phobos}-2 and Phobos in the projection onto the Mars
ecliptic plane $XoY$ on March 22--26, 1989. The $X$-axis points to
the Sun; the $X$-$Y$ plane coincides with the orbital plane of
Mars; the $Y$-axis points in opposite direction of the Mars
orbital velocity; the $Z$-axis is perpendicular to $X$ and $Y$.
The line composed of small squares allocates the part of the
spacecraft trajectory on the circular orbit around Mars where the
magnetic field enhancements were observed.
\end{minipage}
\vspace{8mm}

 The morphology of the magnetic field signatures
caused by the interaction of the Phobos magnetic field with the
solar wind plasma and observed during the time interval of March
22--26, 1989 are presented in Figs.~2--4. The top panels of
Figs.~2--4 present the segment of the spacecraft trajectory in the
projection onto the Mars ecliptic plane $XoY$ where these magnetic
fields were acquired to demonstrate that the magnetic field
enhancements in the regular part are not associated with Mars.
With this aim, we indicate the location of the Mars bow shock
deduced from the {\it Phobos}-2 data and marked by the solid line
in Fig.~1. The distance from the dayside surface of Mars to the
position of the Mars shock is less then 400--1000 km. Studying the
{\it Phobos}-2 data, Schwingenschuh et al., (1992) noted that the
dependence of the bow shock position in the subsolar region on the
solar wind dynamic pressure is weak. Really, over up month since
February over March 1989 the spacecraft crossed the Mars bow
shocks at $X\simeq0$, $Y=+9600$ and $X\simeq0$, $Y=-9600$ km. The
position of the Mars bow shock crossing was the same during March
22-26, 1989. This location of the Mars bow shock was confirmed by
a large database of bow shock crossings by the MGS spacecraft
(Vignes et al., 2000; Luhmann et al., 2002). So the magnetic field
enhancement in the regular part can not be really attributed to
the Mars. The S/C crossed a stationary slowly moving structure,
which approached to and then moved away from the S/C accordingly
with the scheme of the S/C flight.

The plot of the distance Rx between Phobos and the spacecraft versus
the time of the observation is represented in the bottom panels of
Figs.~2--4 to illustrate that the magnetic field disturbances are
really associated with Phobos. There is a clear correspondence between
the disturbance of the magnetic field and the approaches of the
spacecraft to Phobos. The magnetic field disturbances are associated
with the solar wind interaction with Phobos. Below, the signs of the
magnetic field components ({\bf Bsx, Bsy, Bsz}) of the undisturbed
solar wind are presented to indicate an association of the
phenomenon with the directions of the interplanetary magnetic field.

The plot of the magnitude of the magnetic field ({\bf B}) versus the
time in the middle panels of Figs.~2--4. An absence of data for some time
intervals leads to the fragmentary plot. The three components of the measured
magnetic field are displayed in (Delva et al., 1994; Yeroshenko, 2000).
In Figs.~2--4, the magnitude
of the observed magnetic field is marked by the solar wind parameters
acquired by {\it TAUS} device) (Vs--the solar wind velocity in km/sec,
Ns--the solar wind density in cm$^{-3}$) to illustrate that the
manifestation of the phenomenon depends on the solar wind parameters.
The arrows mark the time when the parameters were acquired.

\vskip 6mm
\psfig {file=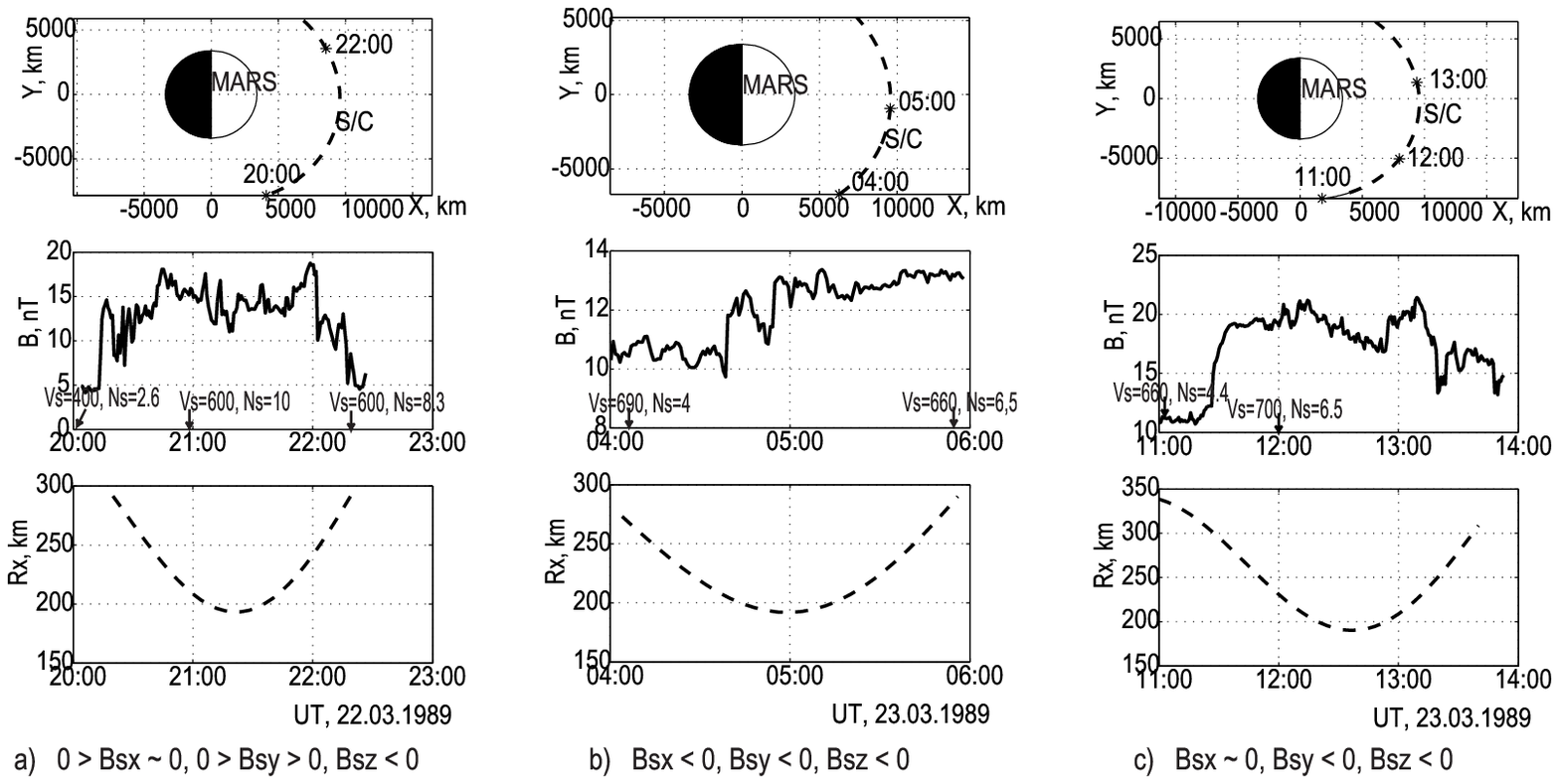}
\vskip 3mm {\small

{\bf Fig. 2.}  {\bf O}n the top the plot of the magnetic field
signatures and the appropriate spacecraft trajectories in the
projection onto the Mars ecliptic plane $XoY$.
{\bf T}he lower graphs correspond to the time history of spacecraft
approaches to the dayside of Phobos inside the unperturbed solar
wind.  {\bf a}) The data from 20:15 to 22:15 on March 22, 1989. {\bf
b}) The data from 04:00 to 06:00 on March 23, 1989. {\bf c}) The data
from 11:00 to 13:45 on March 23, 1989.}

\vskip 10mm
\psfig {file=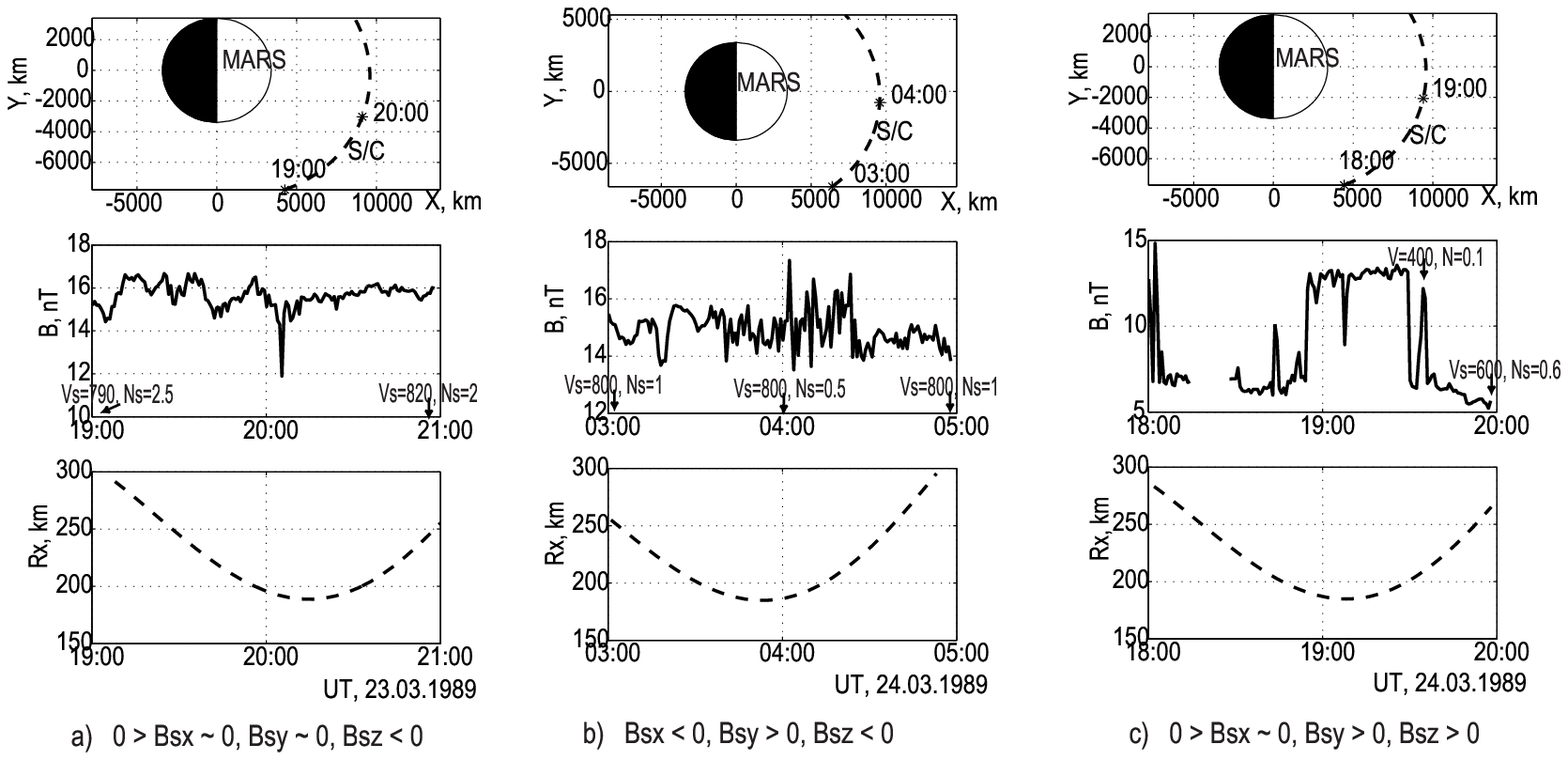}
\vskip 5mm
{\small{\bf Fig. 3.}  {\bf
S}imular to Figure 2, except {\bf a}) The data from 19:00 to 21:00
on March 23, 1989. {\bf b}) The data from 03:00 to 05:00 on March
24, 1989. {\bf c}) The data from 18:00 to 20:00 on March 24,
1989.}
\vskip 10mm

\psfig {file=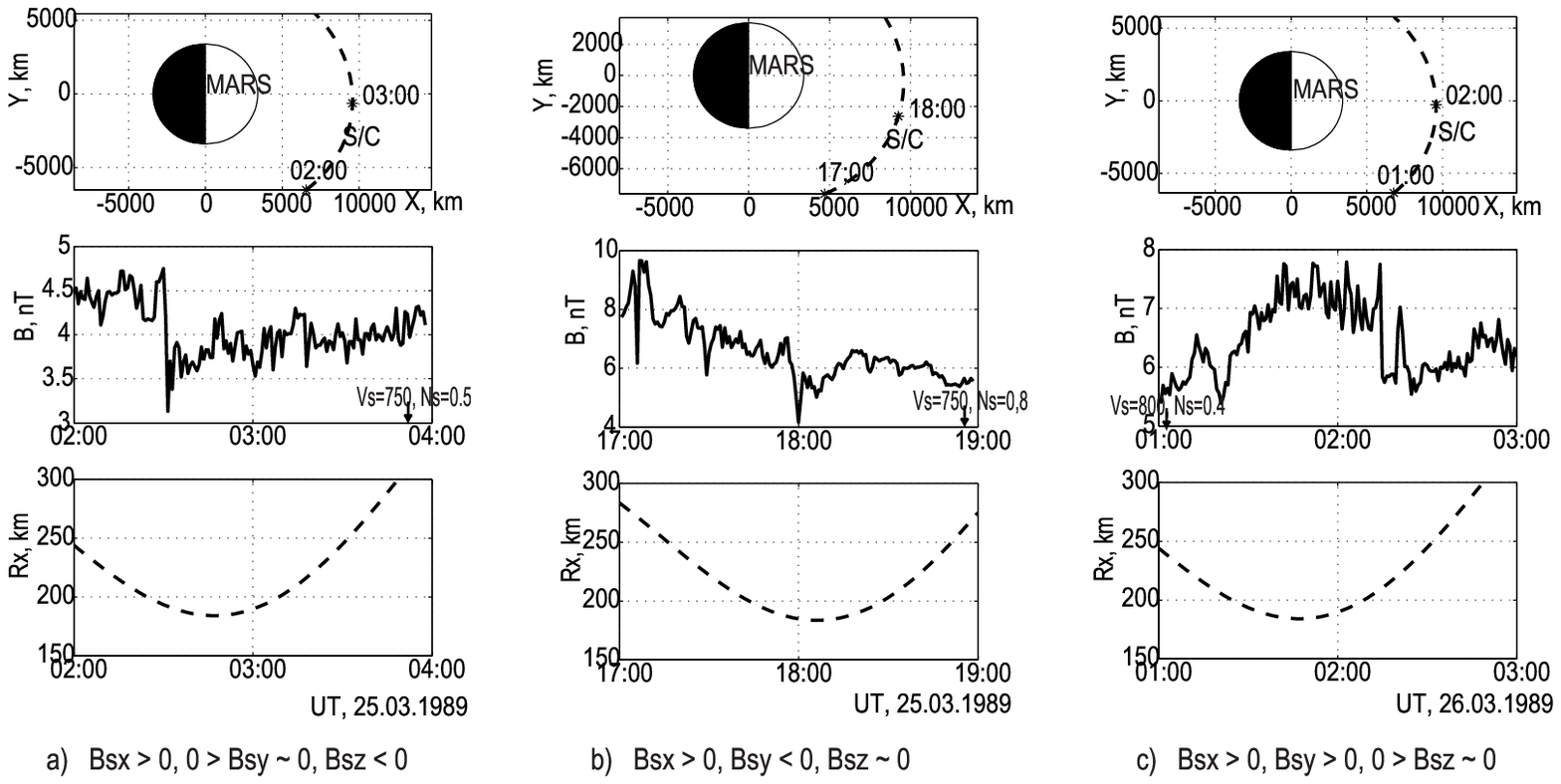} \vskip 5mm {\small{\bf Fig. 4.}  {\bf
S}imular to Figure 2, except {\bf a}) The data from 02:00 to 04:00
on March 25, 1989. {\bf b}) The data from 17:00 to 19:00 on March
25, 1989. {\bf c}) The data from 01:00 to 03:00 on March 26,
1989.}
\vskip 10mm

\section*{Discussion}
The small characteristic size of the Phobos obstacle to the solar wind
have to give specific signatures of the solar wind-asteroid
interaction. The typical free path for solar wind particles is much
greater than the size of Phobos and the ion gyroradius for the solar
wind plasma have the same order as the Phobos size. Therefore, a
magneto-hydrodynamic approximation fails and it is necessary to use a
kinetic theory to study the solar interaction with Phobos and other
small bodies. In addition, it is worth noting that such MHD terms as
``bow shock'' and ``subsolar magnetosheath'' are inappropriate for
description of the solar wind interaction with a small magnetized
asteroid.

The kinetic theory was used for description of interaction of
spacecraft with magnetosphere plasma (Alpert et al., 1964). We can
employ results inferred from the study. According to the kinetic
theory, the density and magnetic field of the plasma pile up in front
of the obstacle to the flow. The pile up becomes significant when the
ion skin depth is comparable to the actual size of the obstacle. The
same conclusion for the interaction of the solar wind with a magnetized
asteroid was inferred by Blanco-Cano et al. (2002), which made the MHD
simulation with ion kinetic effects.

Figures 2--4 give a lesson for the study of the solar wind interaction
with a small, magnetized object. The magnetic field signatures observed
near day-side of Phobos show the response of the solar wind to the
Phobos obstacle. The draping magnetic field around Phobos appears at
distances of 200--300~km from the Phobos day-side, the distance depends
on the solar wind plasma parameters. The events displayed in Fig.~2
result from the interaction of the Phobos obstacle
with the solar wind plasma having high density (Ns) from 3~cm$^{-3}$
up 12 cm$^{-3}$. Figures 3 and 4 give the examples of the interaction
in the plasma with lower density, Ns is between 2~cm$^{-3}$
and 0.4~cm$^{-3}$. The velocity of the solar wind changes slightly
during these observations. Therefore, the density of the solar
wind plasma plays a significant role in formation of the size
and shape of the draping and compressional region near the dayside of
Phobos and around Phobos. The density and magnetic field of the solar
wind plasma pile up in front of the obstacle that Phobos and its
magnetic field represent to the solar wind. The pile up becomes
significant when the proton skin depth is comparable with the actual
size of the Phobos obstacle.

We can take the actual size of the Phobos obstacle from the paper by
Mordovskaya et al. (2001) or estimate it from Figure 6 by calculating
the size of region occupied by the planetary field of Phobos. This size
is about 150--170 km and the relevant density is 1.8--2.3~cm$^{-3}$.
Formation of a shock-like structure upstream of Phobos will take place
for the solar wind plasma with density larger than 1.8--2.3~cm$^{-3}$.
The magnetic field signatures in Fig.~2 demonstrate the shock-like
structure upstream of Phobos and it is seen that in this case the
draping of the field is stronger. Figures 3--4 show an absence of the
shock-like structure ahead of Phobos magnetopause and the week draping
of the field around the Phobos obstacle because the plasma density was
low. For other plasma densities observed, the ion scale
length $l_s$ is $l_s$=93~km for Ns~=~6~cm$^{-3}$; $l_s=130$~km for
Ns~=~3~cm$^{-3}$; $l_s=160$~km for Ns~=~2~cm$^{-3}$; $l_s=294$~km for
Ns~=~0.6~cm$^{-3}$. It easily seen that the nature of the
interaction displayed by the magnetic field signatures in Figs. 2--4 is
related to the ion scale lengths. The presence or absence of a
shock-like structure ahead of the Phobos magnetopause are consistent
with the ratio of the proton skin depth to the actual size of the
Phobos obstacle. Depending on this ratio, some ions and magnetic field
line will pile up in front of the Phobos magnetic barrier, forming
fanciful patterns of the density and magnetic field signatures. The
field signatures observed are various and depend on the density of the
solar wind flow.

It is worth dwell upon an additional feature of behavior of the plasma
density and the magnetic field (see Figs.~3b and 3c). The density value
decreases by a factor of 2 in Fig.~3b and by an order of magnitude with
a closest approach to Phobos, the decrease indicates to the
absence and lack of plasma, at least that of the solar wind, near
regions adjacent to the Phobos magnetopause.

On March 24, 1989, the dynamic pressure of the solar wind begins to
decrease. The density of the solar wind decreased down to 0.5--1
cm$^{-3}$ during this period. When the speed Vs of the solar wind falls
down to 600 km/s, it was possible to observe a remarkable event shown
in Fig.~3c. During 18:43--19:41 on March 24, 1989 the magnetic field
signature has a sharp rise with a characteristic ``magnetopause-like''
behavior demonstrating clear encounter with an intrinsic magnetic field
of Phobos. The field magnitude increased by 80\% with respect to the
background level, while the plasma density value was, apparently, close
to the lower threshold of sensitivity of the plasma detector. In
detail, we consider the observation of the planetary magnetic field of
Phobos in the next section.

The performed analysis of the magnetic field signatures observed near
day-side of Phobos confirms that the nature of the interaction of
Phobos with the solar wind is related to ion scale lengths and
highlights the importance of studying the interaction of the solar wind
with a small magnetized object using the kinetic theory. The small size
of Phobos is the main cause of departures from a planet-like
interaction with the solar wind. In agreement with the kinetic theory,
the magnetic field of the solar wind drapes around the Phobos obstacle,
the effect depends on the ratio of the proton skin depth to the actual
size of the Phobos obstacle. In Figs.~2-4, except for Fig.~3c, the
magnetic field signatures are caused by the interaction of the solar
wind with Phobos. On the other hand, Fig.~3c shows the direct
measurements of the planetary magnetic field of Phobos.

\section*{THE OBSERVATION OF THE PLANETARY MAGNETIC FIELD OF PHOBOS}

The periapsis of the {\it Phobos}-2 spacecraft was far from the dayside
surface of Phobos, the smallest altitude is 170--180 km. We could
directly probe the regions to see pure planetary magnetic field when
the dynamic pressure of the solar wind dropped. There is a unique
case of the manifestation of the Phobos intrinsic magnetic field during
18:43--19:41 on March 24, 1989 (Fig.~3c). In fig. 5 the three components
of the measured magnetic field, velocity V, and density N of plasma
for 18:00--20:00 of March 24, 1989 are displayed.
\vskip 5mm
\psfig {file=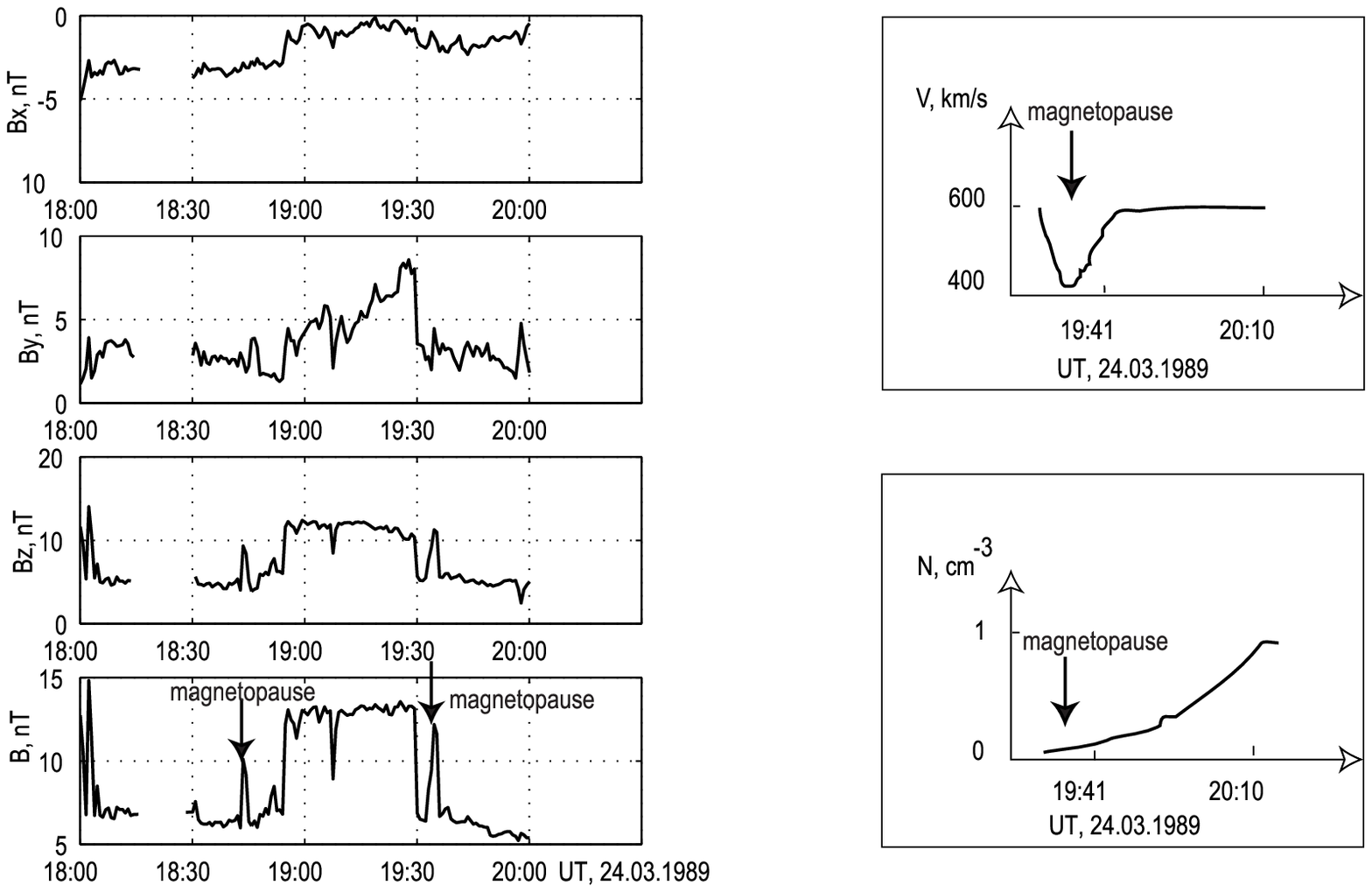}
\vskip 5mm {\small {\bf Fig. 5.}  {\bf M}agnetic field and plasma
data illustrating the encounter with a planetary magnetic field of
Phobos as delineated by the arrows. {\bf L}eft panels: plot of the
magnetic field components ($B_{x}, B_{y}, B_{z}, B$) in nT. {\bf
R}ight panels: plots of the plasma speed {\bf V} in km/s and the
plasma density {\bf N} in $ cm^{-3} $}

\vskip 5mm

The data show the sharp boundary that clearly separates different
plasma regions. The magnetic field enhancement (B) is an almost
rectangular pulse of a symmetrical shape with a characteristic
``magnetopause-like'' behavior. The encounter into the region
occupied by the magnetic field of Phobos is indicated by a
magnetopause crossing. The velocity signature says in favour of
the existence of the magnetopause. Here we say about the
magnetopause as a boundary separating the regions of the planetary
magnetic field of Phobos and the solar wind plasma. The plasma
density value was of order of 0.1~cm$^{-3}$, apparently, close to
the lower threshold of sensitivity of the plasma detector. The
absence of the solar wind plasma there gives a strong argument in
favour of the existence of the region occupied by of the Phobos
magnetic field.

The satellite seems to be in the subsolar point of the magnetosphere of
Phobos near the magnetopause. The estimation of the magnetic moment of
Phobos $M^{\prime}$ in the dipole approximation with help of the
equation of pressure balance for the solar wind and the magnetic field
of Phobos at the magnetopause and the experimental data was given
by Mordovskaya et al. (2001):
$2N_{s}m_{p}V^{2}_{s}=1/8\pi\left(2M^{\prime}/D^{3}\right)^2$, where
$D$ is the distance from the center of the planet up to its subsolar
point, N$_{s}$ and V$_{s}$ are the density and the speed of the solar
wind, respectively, $m_p$ is the proton mass.  The measured values of
the concentration and the speed of the solar wind, which are used to
estimate the magnetic moment $M^{\prime}$, are N$_{s}$=0.17 cm$^{-3}$
and V$_{s}$=617 km/s. The estimate obtained is $M'\simeq10^{15}$
A$\cdot$m$^2$. It is a source with such magnetic moment in Phobos that
can deflect the solar wind flow at the distance over 150--170 km from
the dayside of Phobos.
\begin{minipage}{85mm}
\rightskip 5mm
\psfig {file=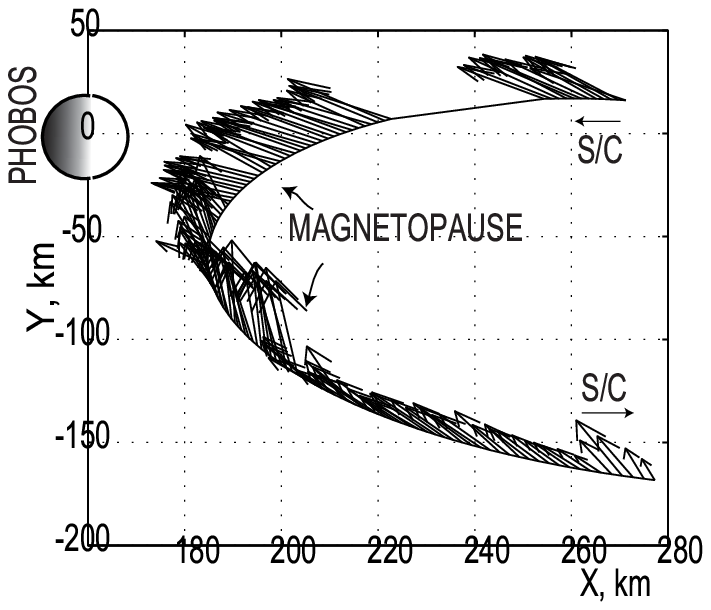}
{\small {\bf Fig. 6.}  {\bf P}lot the amplitudes, the
direction of vectors of the magnetic field Bx/By at 45-s intervals
along the S/C trajectory in the Phobos centric  coordinate system.
The data are from 18:00 to 20:00 on March 24 , 1989.}
\end{minipage}
\begin{minipage}{93mm}
\vskip 2mm
To give an additional grasp of the phenomenon, we
display the {\it Phobos}-2 encounter with the planetary magnetic
field of Phobos with the help of magnetic field lines in Fig.~6.
The amplitudes, the direction of vectors of the magnetic field
{\bf Bx/By}, and the S/C trajectory in the projection onto the
Mars ecliptic plane $XoY$ are given in the Phobos centric
coordinate system. The field observed along the trajectory is
represented by a scaled vector projection of {\bf B} originating
from the position of the spacecraft at the corresponding times,
the time interval between two consecutive measurements was 45 s.
Analyzing the magnetic field direction near Phobos, we can
determine with a great accuracy the boundary between completely
open field lines of the solar wind and those with at least one end
in Phobos. A dramatic change in the field line topology from 18:43
to 19:41 on March 24, 1989 indicates a transition from field lines
with no connection to Phobos to field lines with at least one end
in Phobos.
\end{minipage}

\section*{CONCLUSION}

In this study we have considered the solar wind interaction with Phobos
using the magnetic field and plasma data from the {\it Phobos}-2
mission during the closest fly-by on March 22--26, 1989. The magnetic
field disturbances observed are correlated with the approaches of the
spacecraft to Phobos. In the magnetic field signatures the
manifestation of the interaction depends on the plasma parameters of
the solar wind and are different for low and high plasma density. The
draping magnetic field around Phobos appears at distances of
200--300~km from the Phobos day-side due to the density and magnetic
field pile up in front of the Phobos obstacle. The nature of the
interaction and the magnetic field signatures observed are consistent
with the ratio of the proton skin depth to the actual size of the
Phobos obstacle to the solar wind. On the other hand, the clear
observation of the planetary magnetic field of Phobos is highly
controlled by the ${\bf Bsz,\,\,Bsy}>0$ directions of the
interplanetary magnetic field.

It was shown how Phobos deflects the flow of the solar wind. The
subsolar stand-off distance of the deflection is about 16--17 Phobos
radii. The planetary magnetic field of Phobos was clearly observed
during 18:43 to 19:41 on March 24, 1989 at 170 km from the dayside of
Phobos. The plasma and magnetic field data confirm this fact.
By using the equation of pressure balance for the solar wind and the
magnetic field of Phobos at the magnetopause, we calculated the
equivalent magnetic moment $M'$, source of which in Phobos leads to the
development of such an obstacle for solar wind flow around Phobos. In
the dipole approximation, the magnetic moment of Phobos is
$M'\simeq10^{15}$ A$\cdot$m$^2$.

It is worth noting that the sizes of Phobos and an ion gyroradius of
the interplanetary plasma have the same order of magnitude. This fact
point out the Phobos magnetosphere to be a system having fundamental
differences with that of the Earth.

The cause of density pile up is related to the actual size of the
obstacle that Phobos and its magnetic field represent to the solar wind
and highlights the importance of studying the interaction of the solar
wind with a small magnetized object with the use of kinetic theory.

In conclusion, the magnetization of Phobos substance is 0.15 CGS. The
third part of volume of Phobos should consist of a magnetic substance
similar to a magnetite Fe$_3$O$_4$ in order to obtain the given
magnetization of Phobos. Since the density of Phobos is about 2
g/cm$^3$, we can suggest two explanations for the magnetization
observed. First, Phobos is non-uniform and there exists an immense
piece of a magnetic material within it. Second, Phobos consists of
small pieces of a magnetic substance immersed into a non-magnetic low
density material.

\section*{ACKNOWLEDGEMENTS}

We thank G.~Kotova for providing some unpublished plasma data
(experiment TAUS).
V. Mordovskaya appreciates helpful communications and
assistance of Ye. G. Yeroshenko.

\end{document}